\documentclass[conference]{IEEEtran}

\usepackage{cite}
\usepackage{amsmath,amssymb,amsfonts}
\usepackage{graphicx}
\usepackage{xcolor}
\usepackage[whole]{bxcjkjatype}
\usepackage{xurl}
\usepackage{booktabs}
\usepackage{colortbl}
\usepackage{eso-pic}

\AddToShipoutPictureBG*{
  \AtPageUpperLeft{
    \raisebox{-1.2cm}{\hspace{2cm}\fbox{\parbox{0.95\textwidth}{\small This paper was originally presented at IEEE CCNC 2025. An extended version of this work has been published in IEEE Access.\\
    For citation, please refer to the IEEE Access version: D.~Chiba, H.~Nakano, and T.~Koide, ``DomainHarvester: Uncovering Trustworthy Domains Beyond Popularity Rankings,'' IEEE Access, 2025, https://doi.org/10.1109/ACCESS.2025.3539882.}}}
  }
  \AtPageLowerLeft{
    \raisebox{1.2cm}{\hspace{2cm}\parbox{0.95\textwidth}{\footnotesize © 2025 IEEE. Personal use of this material is permitted. Permission from IEEE must be obtained for all other uses, in any current or future media, including reprinting/republishing this material for advertising or promotional purposes, creating new collective works, for resale or redistribution to servers or lists, or reuse of any copyrighted component of this work in other works.}}
  }
}

\begin{document}

\title{DomainHarvester: Harvesting Infrequently Visited Yet Trustworthy Domain Names}

\author{
\IEEEauthorblockN{Daiki Chiba, Hiroki Nakano, and Takashi Koide}
\IEEEauthorblockA{NTT Security Holdings Corporation \& NTT Corporation, Tokyo, Japan \\ Email: daiki.chiba@ieee.org}
}

\maketitle

\begin{abstract}
In cybersecurity, allow lists play a crucial role in distinguishing safe websites from potential threats. Conventional methods for compiling allow lists, focusing heavily on website popularity, often overlook infrequently visited legitimate domains. This paper introduces DomainHarvester, a system aimed at generating allow lists that include trustworthy yet infrequently visited domains. By adopting an innovative bottom-up methodology that leverages the web's hyperlink structure, DomainHarvester identifies legitimate yet underrepresented domains. The system uses seed URLs to gather domain names, employing machine learning with a Transformer-based approach to assess their trustworthiness. DomainHarvester has developed two distinct allow lists: one with a global focus and another emphasizing local relevance. Compared to six existing top lists, DomainHarvester's allow lists show minimal overlaps, 4\% globally and 0.1\% locally, while significantly reducing the risk of including malicious domains, thereby enhancing security. The contributions of this research are substantial, illuminating the overlooked aspect of trustworthy yet underrepresented domains and introducing DomainHarvester, a system that goes beyond traditional popularity-based metrics. Our methodology enhances the inclusivity and precision of allow lists, offering significant advantages to users and businesses worldwide, especially in non-English speaking regions.
\end{abstract}

\begin{IEEEkeywords}
Domain Name, Top List, Allow List
\end{IEEEkeywords}

\section{Introduction}
\label{sec:introduction}
In cybersecurity, the deployment of allow lists, previously referred to as whitelists~\cite{ncsc_terminology}, plays a crucial role.
These lists are pivotal in reducing false positives, where legitimate websites are incorrectly identified as malicious threats.
The expansive and evolving nature of the web intensifies the challenge of manually curating comprehensive allow lists.
This research addresses the critical question: How can allow lists be curated systematically and dynamically to accurately include legitimate, yet less-visited websites, while minimizing the inclusion of malicious domains?

An allow list consists of a rigorously vetted set of websites and domains considered safe based on strict criteria, thus exempting them from being marked as malicious.
Traditionally, allow lists have been compiled from top lists, based on the premise that a website's popularity indicates its legitimacy, and have found widespread use in both commercial and academic contexts.
These lists are compiled by assessing web traffic through specific metrics, creating a hierarchy purportedly indicative of trustworthiness.

Nevertheless, top lists have a critical shortcoming: they often neglect legitimate but less-visited or niche websites.
This oversight disproportionately impacts sites related to emerging businesses, new services, or those serving specific regions.
An empirical study revealed biases in top lists, particularly the underrepresentation of non-English-speaking regions, underscoring the issue's global magnitude~\cite{DBLP:conf/imc/Ruth0WVD22}.

Omitting legitimate sites from allow lists leads to more false positives in detecting malicious sites.
Harmless sites are mistakenly identified as threats, making it harder to distinguish between authentic sites and phishing schemes.
This issue stems from traditional allow lists' reliance on top lists, often overlooking niche yet legitimate websites.
As a result, security mechanisms, including DNS, proxies, and antivirus software, may inadvertently restrict access to legitimate domains.
This inability to accurately separate genuine sites from malicious counterparts poses significant challenges for non-English-speaking communities, underscoring the need for a more inclusive, dynamic approach to allow list generation.

In response, we propose a novel, bottom-up methodology that deviates from the conventional top-down, popularity-driven framework.
Our method capitalizes on the web's inherent hyperlink structure, using hyperlink connections and network foundations to discover legitimate but less-visited sites.

DomainHarvester, our proposed system, is designed for the continuous generation of allow lists that encompass legitimate websites, including those less-visited.
Starting with seed URLs, DomainHarvester conducts regular web and DNS explorations to collect domain names linked to these seeds.
Utilizing a machine learning (ML) model, informed by web content and DNS records, combined with a Transformer-based language model (LM), it assembles a roster of trustworthy domains.

Operating DomainHarvester on a weekly basis generates two distinct allow lists: one with a global focus and another oriented towards local preferences.
Our analysis, compared with six existing top lists, showed that DomainHarvester's allow lists encompass a wider array of less-visited domains, with overlaps of only 4\% globally and 0.1\% locally.
Additionally, it substantially reduces the likelihood of including malicious domains, enhancing the security and reliability of the generated allow lists.

\section{Background and Related Work}
\label{sec:background}

\subsection{Characteristics of Top Lists}
\label{sec:what-is-top-list}
Top lists rank websites or domains based on popularity, employing various criteria and methodologies.
The data used, the definition of popularity, and the aggregation methods vary significantly among different top lists.
We introduce six typical top lists widely used in prior research.

\noindent\textbf{Alexa Top 1 Million} was widely recognized and utilized in the academic community~\cite{DBLP:conf/imc/Ruth0WVD22}.
This list ranked websites based on their average daily visits and page views.
However, the service was discontinued in May 2022.

\noindent\textbf{Cisco Umbrella 1 Million} is a list ranking the most popular domains based on DNS queries within Cisco's DNS service.

\noindent\textbf{Majestic Million} ranks websites' popularity based on web links, as provided by Majestic SEO.

\noindent\textbf{Tranco Top Million List} is a top list created based on a method proposed in a 2019 research paper~\cite{DBLP:conf/ndss/PochatGTKJ19}.
It combines the three lists (Alexa, Umbrella, and Majestic) and has been frequently cited in academic research.

\noindent\textbf{Secrank List} is a top list developed using a novel method proposed in a research paper~\cite{DBLP:conf/uss/XieTZLLD022}, which utilizes a specific voting algorithm based on the number of DNS requests for each domain.

\noindent\textbf{Chrome User Experience Report (CrUX)} ranks the popularity of websites using data from Google Chrome users.

\subsection{Using Top Lists as Allow Lists}
\label{sec:top-list-based-allow-lists}
Under the assumption that popularity suggests safety, top lists have often been used as de facto allow lists---compilations of domains considered trustworthy or benign.
To demonstrate how top lists are used as allow lists, we present two key examples.

\noindent\textbf{False Positive Mitigation.}
While the internal logic of most security services is not publicly disclosed, a case study indicated that top lists are used to reduce false positives.
For instance, Quad9 is believed to include domains from Majestic's top list in its allow list~\cite{DBLP:conf/ndss/PochatGTKJ19,DBLP:conf/uss/XieTZLLD022}.
A commercial product designed to block malicious DNS requests also adopted Alexa Top 1M to prevent future false positives~\cite{dnswatch}.

\noindent\textbf{Benign Labeling.}
Top lists have been widely used to label domains as benign for machine learning model training and refining target data in systems and analysis processes~\cite{DBLP:conf/uss/AntonakakisPDLF10,DBLP:conf/dsn/ChibaYASYMG16,DBLP:journals/compsec/ChibaAYHMG18}.
When detecting phishing sites and squatting domains, higher-ranked domains in top lists are often assumed to be the genuine targets of imitation~\cite{DBLP:conf/ndss/AgtenJPN15,DBLP:conf/raid/0001HKSGA19,DBLP:conf/imc/Suzuki0YMG19}.

\section{Proposed System: DomainHarvester}
\label{sec:proposed-system}

\subsection{Design Goals}
\label{sec:design-goals}
We propose DomainHarvester, a system designed to generate a list of infrequently visited yet trustworthy web domains.
The system aims to achieve the following two goals:

\noindent\textbf{Goal 1: Incorporate Unpopular Trustworthy Domains.}
Unlike top lists, our objective is to enhance coverage and diversity by incorporating more infrequently visited yet trustworthy domains.
Our definition of \textit{trustworthy} is not based on the popularity metrics of top lists but refers to sites that are explicitly linked from a trustworthy site, are actively managed, and do not engage in obviously malicious activities.

\noindent\textbf{Goal 2: Maintain High Variability.}
Our goal diverges from the stability sought by top lists~\cite{DBLP:conf/uss/XieTZLLD022}.
We aim to maintain high variability, allowing for the rapid inclusion of newly discovered unpopular yet trustworthy domains and the prompt removal of domains that become untrustworthy.

\begin{figure}[!t]
  \centering
  \includegraphics[width=\linewidth]{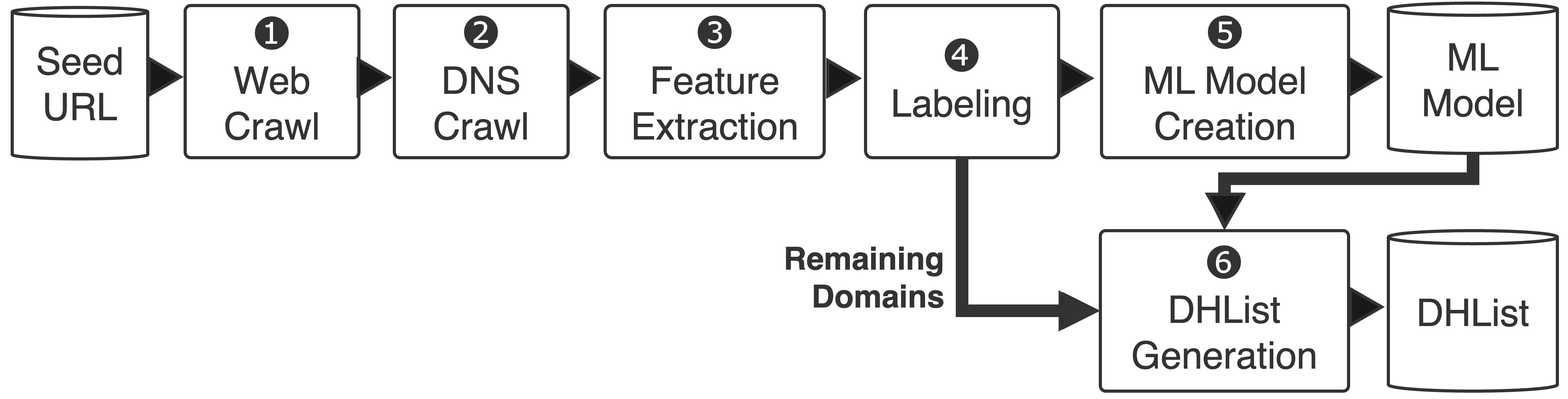}
  \caption{DomainHarvester System Overview}
  \label{fig:system-overview}
\end{figure}

\subsection{System Overview}
\label{sec:system-overview}
To achieve the aforementioned design goals, we introduce DomainHarvester.
Figure~\ref{fig:system-overview} provides an overview of the system.
DomainHarvester periodically performs web crawling on seed URLs to discover domain names linked from those websites and collects their web content (❶).
It also conducts DNS crawling for each discovered domain name to extract DNS record information at that point in time (❷).
Following this, features are derived from the web and DNS crawling results for that iteration (❸).

Using the differences between the current and previous iterations of web and DNS crawling, domains considered to be untrustworthy are identified accordingly (❹).
This step requires data from the two most recent iterations and is executed from the second periodic execution onwards.

Subsequently, a supervised machine learning model is trained to identify such untrustworthy domain names (❺).
Finally, domains not labeled as untrustworthy in step ❹ are processed using the supervised machine learning model to filter out domains estimated to be untrustworthy.
The remaining domain names are compiled into a list of unpopular yet trustworthy domains, termed the DomainHarvester List (DHList), thus fulfilling Design Goal 1 (❻).

The periodic execution contributes to Design Goal 2's emphasis on high variability.
Ideally, changes to the seed URLs' websites and DNS records would be monitored frequently to reflect updates.
However, ethical considerations in this research limit the frequency of access to the seed URLs.
While the frequency of periodic execution is set to weekly for the duration of this study, it may be adjusted in the future to better suit operational needs or to reflect changes in the web's dynamics.
For ethical considerations related to the frequency of access and data collection, please refer to Section~\ref{sec:ethical} for a detailed discussion.

Each step is detailed further in the subsequent sections.

\subsection{❶ Web Crawl}
\label{sec:web-crawl}
In this step, we conduct a web crawl weekly on selected seed URLs.
The underlying hypothesis is that domain names linked from reliable seed URLs are likely to be trustworthy, given the web's link structure.
The seed URL serves as the starting point for web crawling, and the trustworthiness of linked domains is assessed in subsequent steps.
The specific seed URLs evaluated in this study are detailed in Section~\ref{sec:experimental-setup}.

During the crawl, we extract links from websites associated with seed URLs, along with server certificate information and web content.
Each extracted link is then processed according to its Pay-level Domain (PLD).
In this study, the term PLD is used to refer to the smallest unit of a domain that a user can directly register.
For example, the PLD for \texttt{www.example.com} is \texttt{example.com}, and for \texttt{www.example.co.uk}, it is \texttt{example.co.uk}.
PLDs allow us to treat web pages at the same user-registrable domain level, even if the Top-Level Domains (TLDs) differ.

The processing is as follows:

\noindent(1) If the PLD of the link destination matches the PLD of the seed, the crawling process continues up to a pre-specified depth, with web access performed again for that link destination to extract further links.

\noindent(2) If the PLD of the link destination does not match the PLD of the seed, web access is performed only once for that link destination to collect the content, and the extraction process for that branch concludes.

\begin{figure}[!t]
  \centering
  \includegraphics[width=\linewidth]{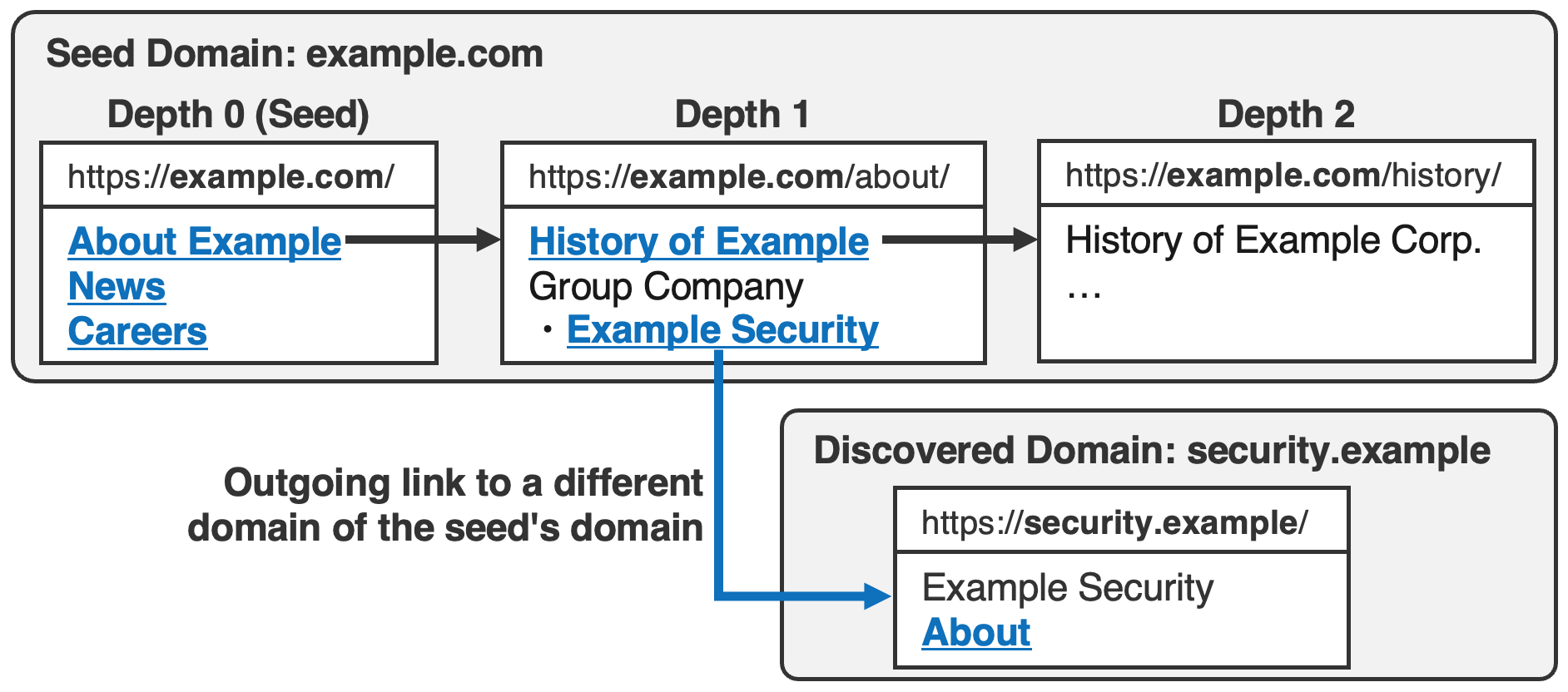}
  \caption{Web Crawling Process Diagram}
  \label{fig:web-crawl}
\end{figure}

Figure~\ref{fig:web-crawl} provides an overview of the web pages that are crawled and their processing.
For instance, the seed \texttt{https://example.com/} represents Depth 0.
The link \texttt{https://example.com/about/}, which shares the same PLD at Depth 1, is targeted for further link extraction.
Conversely, the external link \texttt{https://security.example/}, which has a different PLD, is accessed only once.

To ensure the ethical conduct of our web crawling activities, we adhere to several principles.
We maintain a minimum interval of 3 seconds between server accesses and respect robots.txt files. This ensures that our activities do not negatively impact the performance or availability of the target websites.

In this study, the maximum crawling depth is set to 3.
The decision to set the maximum crawling depth to 3 was made after considering the balance between obtaining a comprehensive overview of linked domains and minimizing the impact on web server resources.
This depth allows us to explore domains linked directly and indirectly to the seed URL while ensuring that our crawling remains focused and relevant to our research objectives.

\subsection{❷ DNS Crawl}
\label{sec:dns-crawl}
We perform a DNS crawl weekly on all domains discovered through the web crawl.
For each domain, we collect five types of DNS records: NS, A, AAAA, MX, and TXT.
The collection of these DNS records allows DomainHarvester to evaluate the infrastructure and security stance of each domain, contributing to a more accurate evaluation of their trustworthiness.
If an A or AAAA record is found, the corresponding country and organization are identified using the GeoIP Database.

In addition, we evaluate each domain's adherence to seven DNS security mechanisms~\cite{DBLP:conf/globecom/Yajima0YM21} (DNSSEC, Certification Authority Authorization (CAA), Sender Policy Framework (SPF), DomainKeys Identified Mail (DKIM), Domain-based Message Authentication, Reporting, and Conformance (DMARC), Mail Transfer Agent Strict Transport Security (MTA-STS), and DNS-based authentication of named entities (DANE)).
The adherence to these DNS security mechanisms is a strong indicator of a domain's commitment to security and reliability, making it an essential factor in the trustworthiness evaluation process.

\begin{figure}[!t]
  \centering
  \includegraphics[width=\linewidth]{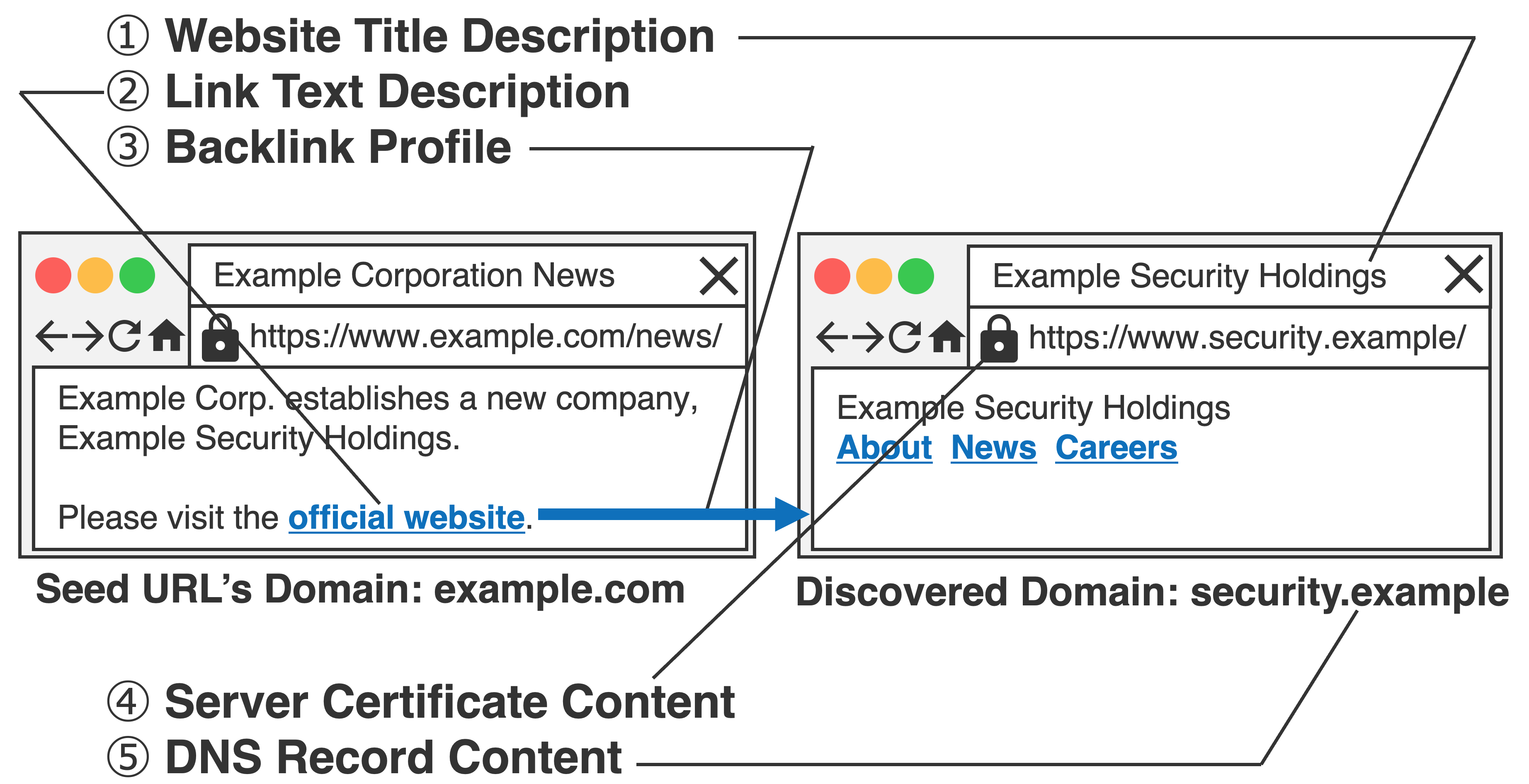}
  \caption{Schematic Representation of Feature Extraction}
  \label{fig:ml-model}
\end{figure}

\begin{table*}
  \centering
  \caption{Summary of extracted features, dimensions, example vectors, and their importance for the machine learning model.}
  \label{table:list-features}
  \centering
  \scriptsize
  \tabcolsep=0.9mm
  {\renewcommand\arraystretch{1.1}
  \begin{tabular}{lrlrlrr}
  \toprule
    &  &  &  &  & \multicolumn{2}{l}{Importance (Section~\ref{sec:characteristics-ml})}\\
    & No. & Feature & Dimension & Example Vectors Corresponding to Figure~\ref{fig:ml-model} & DHList-Global & DHList-Japan \\
  \midrule
  ① Title & 1 & Website Title Description & 768  & title\_0=-0.07,  ... , title\_767=0.04 & 1,489 & 1,717 \\
  \midrule
  ② Link Text & 2 & Link Text Description & 768  & linktext\_0=-0.18, ... , linktext\_767=0.13 & 1,218 & 1,233 \\
  \midrule
  ③ Backlinks & 3 & Backlink Profile & 3,735  & example.com=1, example.org=0, ... & 27 & 24 \\
  \midrule
  ④ Certificate & 4 & Issuer's Organization & 269  & ExampleCert=1, ExampleSign=0, ... & 138 & 215 \\
    & 5 & Issuer's CountryName & 62  & US=1, JP=0, … & 89 & 135 \\
    & 6 & Subject's Organization & 14,090  & ExampleSecurity=1, ExampleCorp=0, … & 92 & 203 \\
    & 7 & Subject's CountryName & 161  & US=1, JP=0, … & 53 & 72 \\
  \midrule
  ⑤ DNS & 8 & \#Records per DNS Record Type & 5  & NS=4, A=2, AAAA=2, MX=6, TXT=7 & 127 & 69 \\
    & 9 & Country corresponding to IPv4 address in the A record & 172  & US=1, JP=0, … & 326 & 93 \\
    & 10 & Organization corresponding to IPv4 address in the A record & 6,497  & ExampleSecurity=1, ExampleCloud=0, … & 258 & 88 \\
    & 11 & Country corresponding to IPv6 address in the AAAA record & 67  & US=1, JP=0, … & 33 & 0 \\
    & 12 & Organization corresponding to IPv6 address in the AAAA record & 772  & ExampleSecurity=1, ExampleCloud=0, … & 32 & 0 \\
    & 13 & Whether DNS security mechanisms are supported & 7  & DNSSEC=0, ... , DANE=0 & 38 & 71 \\
  \bottomrule
  \end{tabular}
  }
\end{table*}

\subsection{❸ Feature Extraction}
In this step, we extract feature values from the five categories shown in Figure~\ref{fig:ml-model}, based on the latest web and DNS crawl results (❶ and ❷).

\noindent\textbf{① Website Title Description.}
This feature aims to recognize the natural language context of a website's title within the discovered domain, helping to identify domains with similar contexts.
We use a multilingual Language Model (LM) from Sentence BERT.
By inputting concatenated titles of all websites within the target domain observed during a specific seed list crawl into the Sentence BERT model, we generate a 768-dimensional feature vector, which is independent of the title's word count.
Sentence BERT is a LM based on the Transformer architecture that provides embedding vectors considering sentence relationships, showing high accuracy in various natural language processing tasks~\cite{DBLP:conf/emnlp/ReimersG19}.

\noindent\textbf{② Link Text Description.}
To identify the context of the text used as the link pointing to the discovered domain, we obtain a feature vector using Sentence BERT, similar to ①.
The feature vector is derived from the concatenated link text in the seed domain that links to the target domain, providing an additional 768 dimensions.

\noindent\textbf{③ Backlink Profile.}
Feature vectors are constructed to capture the number of backlinks to identify the domains that link to the target domain's website.
Backlinks refer to websites that contain links to the domain in question.
For instance, if \texttt{example.com} links to \texttt{example.org}, then \texttt{example.org} has a backlink from \texttt{example.com}.
For instance, if there is one backlink from \texttt{example.com}, we create a feature vector indicating ``backlink from \texttt{example.com}=1'' and ``backlinks from other domains=0.''

\noindent\textbf{④ Server Certificate Content.}
Feature vectors are generated to recognize and identify settings within the certificate's issuer and subject, specifically the Organization and CountryName.

\noindent\textbf{⑤ DNS Record Content.}
We assess the presence of five types of DNS records (NS, A, AAAA, MX, and TXT) collected in the DNS crawl (❷).
If present, the number of those records is extracted as a feature.
Feature vectors are also created to identify the country and organization settings in the A and AAAA records.
Additionally, we extract feature vectors to evaluate each domain's compliance with the seven DNS security mechanisms.
We avoid converting the IP addresses from A and AAAA records into feature vectors to prevent high-dimensional sparse features and overfitting.

Table~\ref{table:list-features} summarizes the features detailed in Figure~\ref{fig:ml-model}, providing a comprehensive overview.
The table includes the features, the number of dimensions for each feature, and a representation of the obtained feature vectors.
It is important to note that the dimensions may vary depending on the data collected in each crawl cycle, noting that the reported maximum value pertains to this study.
Furthermore, the ML model was constructed with careful feature selection and dimensionality reduction based on feature importance to prevent overfitting and optimize training time.

\subsection{❹ Labeling}
\label{sec:labeling}
This step involves labeling domain names considered untrustworthy based on the latest two sets of web crawl results (❶) and DNS crawl results (❷).
The aim of DomainHarvester, as stated in Goal 1 of Section~\ref{sec:design-goals}, is to discover reliable domain names through methods other than the top list approach.
Our hypothesis is that domain names explicitly and intentionally linked from trusted seed URLs, which are also actively managed and not used for obvious malicious activities, are likely to be trustworthy.
Since we have already identified domain names linked from seed URLs in previous steps, this step focuses on labeling by examining the \textit{changes} in the two sets of web and DNS crawl results to determine whether there have been any unintended changes indicative of malicious activities.
We sequentially examine changes across five categories, labeling a domain as untrustworthy if it meets the conditions in any category, thereby concluding the labeling process.

\noindent\textbf{DNS Record Changes.}
By comparing the results of the latest two DNS crawls, we label domain names as untrustworthy if there has been a change or deletion (NXDOMAIN) in the combination of Name Server (NS) records for DNS.
Such changes indicate that the server specified for the domain records has been altered.
While there could be legitimate reasons for intentional changes, the goal of DomainHarvester is to identify domain names that can be used as an allow list; thus, we proactively exclude any that exhibit even potential variability in trustworthiness.

\noindent\textbf{Server Certificate Changes.}
Using the results from the latest two web crawls, we label domain names as untrustworthy if they were extracted as HTTPS links and if an expired certificate was used at any point.
The occurrence of an expired certificate (even if later updated) suggests inadequate management of the domain name or website and indicates a potential for future neglect.

\noindent\textbf{Access Status Changes.}
Utilizing the results from the two most recent web crawls, we label domain names as untrustworthy if they could not be accessed or retrieve a web page due to HTTPS or HTTP connection errors at any point.
While we do not dismiss the possibility that the website may temporarily be under maintenance, the aim of DomainHarvester is to identify highly reliable domain names that can be used as an allow list; therefore, we proactively exclude any domains that exhibit even potential variability in reliability.

\noindent\textbf{Web Backlink Changes.}
Using the results from the latest two web crawls for the same seed URL list composed of multiple seed URLs, we label domain names as untrustworthy if there has been a significant decrease in the backlinks to a domain name from the previous to the current crawl.
By examining the web crawl results from the same seed URL list and their differences, we can identify domain names that have suffered a reduction in backlinks beyond a certain ratio.
This ratio is an adjustable parameter, but for this study, domain names with more than a fifty percent reduction in backlinks are considered untrustworthy.
A decrease in backlinks indicates a deliberate removal of links from certain seed URLs, and we proactively exclude any domains that show even potential variability in trustworthiness.

\noindent\textbf{Domain Parking Changes.}
Using the results from the two most recent DNS and web crawls, we assess whether a domain name is utilizing domain parking.
Domain parking refers to the practice of registering a domain and directing it to a web page containing advertisements and links to other websites, primarily for revenue generation.
Specifically, we identify domains that have ceased domain parking from the previous week to the current week, those that have started using domain parking, and those that continue to use domain parking, labeling any domain names falling into any of these categories as untrustworthy.
The rationale for this labeling is that a domain that has been parked or has shown signs of abandonment is considered untrustworthy, even if it was trustworthy in the past.
To determine whether a domain uses domain parking, we follow the multiple prior studies~\cite{DBLP:journals/compsec/ChibaAYHMG18,DBLP:conf/ndss/AgtenJPN15} and ascertain whether the domain's NS record belongs to known domain parking providers or if the domain name's website exhibits characteristics of domain parking (e.g., display of advertisements only, information on domain name sales).

\subsection{❺ ML Model Creation}
\label{sec:ml-model}
This step involves creating a supervised machine learning model based on the features and labels obtained from the preceding steps.
It is important to note that in the Labeling (❹) step, we have only assigned the label ``untrustworthy'' to certain domains, while all others remain unlabeled or unknown.

We avoid traditional binary labeling of trustworthy/negative and untrustworthy/positive due to its simplistic nature, which does not capture the nuances of domain trustworthiness.
As discussed in Section~\ref{sec:top-list-based-allow-lists}, previous research relying on top lists for trust indicators has led to an overestimation of accuracy by equating domain popularity with reliability, thus overlooking domains that are not well-known but are trustworthy.

To overcome the biases associated with top list-based trust labels and to align with our design goals as described in Section~\ref{sec:design-goals}, we adopt Positive-Unlabeled (PU) learning~\cite{DBLP:conf/kdd/ElkanN08}.
This learning approach is suitable for datasets with positively labeled instances and unlabeled instances, which may include both positive and negative instances.
By using PU learning, which is resistant to top list biases, we can identify domains that may be unpopular but are trustworthy.
Specifically, in this study, we label domain names identified as untrustworthy as positive and randomly sample a subset of the remaining domain names as unlabeled, then apply PU learning.

Although multiple supervised machine learning algorithms are suitable for PU learning, we choose LightGBM due to its high accuracy, rapid learning speeds, and independence from feature scaling.
Hyperparameters are tuned weekly with Optuna to select the optimal settings.
The results of comparisons with other algorithms will be reported in Section~\ref{sec:characteristics-ml}.
The machine learning model is updated regularly based on the results of weekly web and DNS crawls to include the latest data.

\subsection{❻ DHList Generation}
\label{sec:dhlist-generation}
In this step, we generate the DomainHarvester List (DHList), comprising trustworthy domain names.
Specifically, the remaining domains that were not labeled in the Labeling (❹) step are fed into the supervised machine learning model.
The model estimates the probability of each domain name being untrustworthy, excluding those deemed highly likely to be untrustworthy, thereby producing the DHList.
Probability values range from 0.0 to 1.0, with domain names having a probability above a specified threshold being excluded as untrustworthy.
We choose a lower threshold to proactively exclude domains that may be untrustworthy, in line with the objectives of this research.

The number of domains included in the DHList is not fixed, unlike top lists, but varies depending on the domains discovered in a given week.
By using a trained ML model, we estimate and exclude as untrustworthy those domain names that were not explicitly labeled as such but share similar features.
This step excludes domain names already labeled as positive/untrustworthy and used for training the ML model, ensuring that the training and testing datasets remain separate for subsequent evaluations.

\section{Evaluation}
\label{sec:evaluation}
In this section, we evaluate DomainHarvester's effectiveness by addressing three key questions:
Q1. How does DomainHarvester operate on actual seed URLs to generate DHLists?
Q2. What characteristics do the generated DHLists have?
Q3. What characteristics do the machine learning models generated during DomainHarvester exhibit?
To address Q1, we prepared multiple seed URLs, operated DomainHarvester, and analyzed the resulting DHLists.
For Q2, we assessed the characteristics of the domain names in the DHLists.
Lastly, for Q3, we detail the accuracy and feature characteristics of the machine learning models developed by DomainHarvester.

\subsection{Experimental Setup}
\label{sec:experimental-setup}
We created seed lists and set up the operating environment required for running DomainHarvester in production.
While DomainHarvester is designed to accept any seed URL, this evaluation adopts a realistic approach to select a certain number of URLs, primarily from the official corporate websites of publicly listed companies.
These companies are legally obligated to disclose information and generally establish credible websites for investor relations, making them relatively reliable within the web space.
It is important to note that not all websites of publicly listed companies are always trustworthy, as they may be tampered with or inadequately maintained.
However, the seed URLs are only used to initiate web crawling, and the trustworthiness of each domain is determined in later steps, ensuring that untrustworthy domains are not included in the final DHLists.

\begin{table}[!t]
  \centering
  \caption{Top 10 regions in Seed-Global.}
  \label{table:seed-global-region}
  \centering
  \begin{tabular}{rllr}
  \toprule
    & Region & Language & \#Companies\\
  \midrule
  1 & United States & English & 585 (29\%)\\
  2 & China & Chinese & 297 (15\%)\\
  3 & Japan & Japanese & 196 (10\%)\\
  4 & South Korea & Korean & 65 (3\%)\\
  5 & Canada & English/French & 64 (3\%)\\
  6 & United Kingdom & English & 64 (3\%)\\
  7 & India & Hindi/English & 55 (3\%)\\
  8 & France & French & 54 (3\%)\\
  9 & Hong Kong & Chinese/English & 54 (3\%)\\
  10 & Germany & German & 52 (3\%)\\
  \bottomrule
  \end{tabular}
\end{table}

\begin{table}[!t]
  \centering
  \caption{Top 10 sectors in Seed-Japan.}
  \label{table:seed-japan-sector}
  \centering
  \begin{tabular}{rlr}
  \toprule
    & Sector & \#Companies\\
  \midrule
  1 & Information & 549 (15\%)\\
  2 & Services & 510 (14\%)\\
  3 & Retail Trade & 339 (9\%)\\
  4 & Wholesale Trade & 308 (8\%)\\
  5 & Electric Appliances & 241 (6\%)\\
  6 & Machinery & 227 (6\%)\\
  7 & Chemicals & 213 (6\%)\\
  8 & Construction & 156 (4\%)\\
  9 & Real Estate & 139 (4\%)\\
  10 & Foods & 122 (3\%)\\
  \bottomrule
  \end{tabular}
\end{table}

\begin{table*}
  \centering
  \caption{Weekly generation of DHList-Global with seed URLs.}
  \label{table:dhlist-global-result}
  \centering
  \scriptsize
  \tabcolsep=0.7mm
  {\renewcommand\arraystretch{1.1}
  \begin{tabular}{rr|rr|rrrrrr|rr|rrrrr}
  \toprule
    &  & \multicolumn{2}{l|}{❶❷❸} & \multicolumn{6}{l|}{❹ Labeling} & \multicolumn{2}{l|}{❺ ML Model Creation}& \multicolumn{5}{l}{❻ DHList Generation}\\
    &  & \# Seed & \# Discovered & \multicolumn{5}{l}{\# Labeled/Untrustworthy Domains}& Subtotal & \# Unknown & \# Training & \# Input & \# Detected & \# DHList Domains & Diff &  \\
  Week & Date & URLs & Domains (A) & DNS & Cert & Access & Backlink & Parking & (B) & (C) & (D) (=B+C) & (E) (=A-D) & (F) & (G) (=E-F) & (+) & (-) \\
  \midrule
  1 & 2022-11-11 & 2,000  & 103,699  & - & - & - & - & - & - & - & - & - & - & - & - & - \\
  2 & 2022-11-18 & 2,000  & 103,556  & 659  & 685  & 137  & 27  & 128  & 1,636  & 1,636  & 3,272  & 100,284 & 96,067  & 4,217  & 4,217  & 0  \\
  3 & 2022-11-25 & 2,000  & 103,255  & 519  & 671  & 141  & 41  & 126 & 1,498 & 1,498  & 2,996  & 100,259 & 86,619  & \cellcolor{yellow!25}\textbf{13,640}  & 10,992  & 1,569  \\
  \bottomrule
  \end{tabular}
  }
\end{table*}

\begin{table*}
  \centering
  \caption{Weekly generation of DHList-Japan with seed URLs.}
  \label{table:dhlist-japan-result}
  \centering
  \scriptsize
  \tabcolsep=0.7mm
  {\renewcommand\arraystretch{1.1}
  \begin{tabular}{rr|rr|rrrrrr|rr|rrrrr}
  \toprule
    &  & \multicolumn{2}{l|}{❶❷❸} & \multicolumn{6}{l|}{❹ Labeling} & \multicolumn{2}{l|}{❺ ML Model Creation}& \multicolumn{5}{l}{❻ DHList Generation}\\
    &  & \# Seed & \# Discovered & \multicolumn{5}{l}{\# Labeled/Untrustworthy Domains}& Subtotal & \# Unknown & \# Training & \# Input & \# Detected & \# DHList Domains & Diff &  \\
  Week & Date & URLs & Domains (A) & DNS & Cert & Access & Backlink & Parking & (B) & (C) & (D) (=B+C) & (E) (=A-D) & (F) & (G) (=E-F) & (+) & (-) \\
  \midrule
  1 & 2022-09-30 & 3,735  & 95,273  & - & - & - & - & - & - & - & - & - & - & - & - & - \\
  2 & 2022-10-07 & 3,735  & 97,289  & 295  & 1,491  & 144  & 37  & 191 & 2,158 & 2,158  & 4,316  & 92,973 & 89,162  & 3,811  & 3,811 & 0  \\
  3 & 2022-10-14 & 3,735  & 97,542  & 301  & 1,479  & 154  & 17  & 194 & 2,145  & 2,145  & 4,290  & 93,252  & 85,313  & 7,939  & 6,912 & 2,784  \\
  4 & 2022-10-21 & 3,735  & 97,013  & 777  & 1,477  & 137  & 24  & 166 & 2,581  & 2,581 & 5,162  & 91,851 & 87,961  & 3,890  & 2,876 & 6,925  \\
  5 & 2022-10-28 & 3,735  & 97,574  & 311  & 1,465  & 127  & 20  & 203 & 2,126  & 2,126  & 4,252 & 93,322 & 45,669  & 47,653  & 44,726 & 963  \\
  6 & 2022-11-04 & 3,735  & 98,262  & 413  & 1,451  & 134  & 11  & 174 & 2,183 & 2,183  & 4,366  & 93,896 & 70,138  & 23,758  & 6,959 & 30,854  \\
  7 & 2022-11-11 & 3,735  & 97,488  & 230  & 1,436  & 138  & 29  & 171 & 2,004  & 2,004  & 4,008  & 93,480  & 47,095  & 46,385  & 29,409 & 6,782  \\
  8 & 2022-11-18 & 3,735  & 97,866  & 249  & 1,424  & 143  & 19  & 205 & 2,040  & 2,040  & 4,080  & 93,786 & 91,083  & 2,703  & 509 & 44,191  \\
  9 & 2022-11-25 & 3,735  & 98,108  & 339  & 1,195  & 179  & 11  & 201 & 1,925  & 1,925  & 3,850  & 94,258  & 79,200  & \cellcolor{yellow!25}\textbf{15,058}  & 14,656 & 2,301  \\
  \bottomrule
  \end{tabular}
  }
\end{table*}

\noindent\textbf{Seed-Global.}
The initial seed list is derived from the 2022 Forbes Global 2000 edition, published annually by Forbes.
This list ranks the world's largest companies based on sales, net assets, market capitalization, and profits.
This list was chosen as the initial seed because the top pages of these global, large-scale companies are deemed more trustworthy than those of other websites.
As the Forbes Global 2000 list does not include URLs for corporate sites, we manually identified and listed the top-page URLs for all 2,000 companies, based on their names.
Each URL corresponded to a distinct domain.
For e-commerce companies, the corporate homepage was chosen (e.g., \texttt{https://www.aboutamazon.com/}) over the customer-facing top page (e.g., \texttt{https://www.amazon.com/}) as the seed URL.
Table~\ref{table:seed-global-region} shows the regions where the headquarters of the Forbes Global 2000 companies are located, listed in descending order, along with the region's official language.
Although the United States represents 29\% of the total, the list includes a diverse array of companies globally.

\noindent\textbf{Seed-Japan.}
The second seed list is based on companies listed on the Tokyo Stock Exchange as of September 2022.
Similar to the Forbes list, we manually identified 3,735 companies, resulting in an equal number of distinct domains.
This list aimed to capture corporate sites more local-centric than Seed-Global.
Although the local choices included several regions, English-speaking regions with English as the official language (United States, Canada, United Kingdom, India, and Hong Kong) comprised half of the top 10 in the Forbes Global 2000.
We decided to focus on non-English speaking regions.
Japan was selected for two primary reasons: firstly, recent research indicates Japan is undervalued in top lists~\cite{DBLP:conf/imc/Ruth0WVD22}, making it a suitable choice for this study's aim of an allow list independent of top lists.
Second, while China ranks second in Forbes Global, setting up a Microsoft Azure-based crawling environment there requires establishing a local entity, a requirement we could not fulfill.
Therefore, we selected Japan, which ranks third.
Table~\ref{table:seed-japan-sector} displays the sectors covered by the companies in Seed-Japan.

\noindent\textbf{Operating Environment.}
Two DomainHarvester instances were set up in distinct Microsoft Azure regions (US and Japan), facilitating targeted crawling based on regional language settings.
The US-based instance, with English language settings, was used for Seed-Global crawling, while the Japan-based instance, with Japanese language settings, was used for Seed-Japan crawling.
The purpose of this region- and language-specific configuration was to guarantee accurate and contextually relevant data collection from websites predominantly utilized in each respective language.
Each instance was a standard Linux machine with a 6-core CPU, 56GB RAM, and a single GPU (NVIDIA Tesla M60), with static IP addresses corresponding to their regions.
For the DomainHarvester crawling, we implemented software based on Scrapy.
As our primary focus is on legitimate sites, cloaking~\cite{DBLP:conf/sp/ZhangOCSJWSKBWS21}, which is usually associated with malicious sites, does not pose a concern for us.
Although bot detection mechanisms could block automatic scraping, such sites are outside our study's scope.
We chose not to interpret JavaScript for two reasons: to enable lightweight and real-time crawling and to collect explicit external links from the trustworthy seed URLs without capturing links from dynamic content such as advertisements and tracking analytics.
Our crawler identifies itself as such in the User-Agent string, and we do not attempt to disguise it as a non-crawler.

\subsection{DHLists Generation}
\label{sec:dhlists-generation}
Two types of seeds, Seed-Global and Seed-Japan, were prepared for the setup described above, which were subsequently crawled, resulting in two types of DHLists (DHList-Global and DHList-Japan).

\noindent\textbf{DHList-Global.}
This section outlines the outcomes of generating DHList-Global, utilizing Seed-Global as the input seed.
Table~\ref{table:dhlist-global-result} shows the operational week/date for DomainHarvester, the count of URLs in Seed-Global, and the number of domain names discovered through weekly web crawls (❶) (A).
Additionally, the table presents the number of domain names labeled as untrustworthy during the labeling process (❹), broken down into subtotals (B).
It also shows the number of unlabeled (unknown) domain names randomly sampled to match the number of untrustworthy ones (C), and the total number of domain names used as training data (D), which is the sum of B and C.
The input used in DHList Generation (❻) consists of the remaining domain names not used as training data (E), the number of domain names estimated as untrustworthy by the machine learning model creation (F), and the number of domain names output as DHList after excluding those (G).
The far-right column of the table shows the difference in the number of domain names in the DHList compared to the previous cycle (+/-).

As explained in Section~\ref{sec:proposed-system}, DomainHarvester utilizes the differences from the previous cycle.
Therefore, the DHList is generated after the second week (2022-11-18).
The labeling results demonstrate that each category is distinct, with no overlaps, as the categories are determined sequentially.

The results from the machine learning model indicate that a significant number (approximately 90,000) of domains were classified as untrustworthy each time.
Considering that DomainHarvester aims to create an allow list and proactively excludes any domain with even a slight risk (especially since we set the threshold to exclude domain names with a probability value of 0.1 or higher as untrustworthy), these results are justified.
Parameter adjustments could result in fewer domains being identified as untrustworthy and, consequently, more domains remaining on the DHList.

Domains classified as untrustworthy include those associated with past international conferences, temporary promotional sites for movies or events, and neglected official websites or those linked to defunct companies.
Examining the number of DHLists and their differences, we observe that many additions and deletions occur each time, as intended by DomainHarvester.
This satisfies the goal of achieving a high degree of variability (see Section~\ref{sec:design-goals}).
In subsequent sections, DHList-Global will be evaluated for the 13,640 domains generated on 2022-11-25.

\noindent\textbf{DHList-Japan.}
Similarly, we describe the outcome of generating DHList-Japan using Seed-Japan as the seed.
Table~\ref{table:dhlist-japan-result} illustrates the numerical values obtained in generating DHList-Japan, as in the case of DHList-Global.
Overall, as with DHList-Global, many domains are deemed untrustworthy in the machine learning model results, and the number of DHLists varies significantly each time.
The variability noted in DHList-Japan reflects the lower threshold parameters set in the machine learning model, a strategy applied to both DHLists to proactively exclude potentially risky domains, which accounts for the dynamic changes observed.
In subsequent sections, the evaluation of DHList-Japan will focus on the 15,058 domains generated on 2022-11-25.

\subsection{Unpopularity of Domains in DHLists}
\label{sec:unpopularity-dhlists}
This analysis compares the relative unpopularity of domain names in DHLists to those included in top lists.
We collected six existing top lists introduced in Section~\ref{sec:what-is-top-list} on 2022-11-25, along with DHList-Global and DHList-Japan.
To align the collected top six lists with uniform comparison criteria, two normalization processes were applied.
Initially, we converted domains in all lists to the Pay-Level Domain (PLD) format.
Specifically, Umbrella and CrUX lists, which are aggregated by Fully Qualified Domain Names (FQDN) and URLs, respectively, can only be compared after this conversion.
Subsequently, sublists of the top 10,000 domains were generated to serve as an allow list.
We chose this approach for two reasons: first, as discussed in Section~\ref{sec:top-list-based-allow-lists}, subsets of the Top 1M are often utilized rather than the entire list.
Second, the size of our generated DHLists is more comparable to the Top 10k, making it a more appropriate comparison in terms of list size.
Tables~\ref{table:coverage-top10k} and~\ref{table:coverage-top1m} present the coverage results for the six Top 10k and Top 1M lists, respectively, for the domains included in DHList-Global (13,640 domains) and DHList-Japan (15,601 domains).
From these results, we make three observations.
Initially, when compared to the Top 10k lists of similar scale, DHList-Global exhibits at most a 4.19\% overlap, and DHList-Japan has at most a 0.11\% overlap, indicating that DHLists contain almost no popular domains commonly found in the Top 10k lists.
As detailed in Section~\ref{sec:web-crawl}, DomainHarvester follows only explicit links and does not seek domain names associated with automatically loaded content such as advertisements or analytics.
This selective approach contributes to the low overlap, which aligns with our goal of creating an allow list distinct from top lists.
Second, even when compared to the Top 1M lists, which are two orders of magnitude larger than DHLists, DHList-Global has a maximum overlap of only 53.5\%, and DHList-Japan has a maximum of 12.8\%, with DHList-Japan consisting predominantly of more unpopular domain names.
Thirdly, DHLists demonstrate a significantly lower overlap with Secrank and Umbrella, likely due to their DNS-based popularity metrics, as opposed to the other lists (Alexa, Majestic, Tranco, and CrUX), which are primarily based on web-based metrics.
Notably, Secrank's data originates from DNS queries within a single country (China), which may explain the disparities observed with the DHLists.

\begin{table}[!t]
  \centering
  \caption{Overlap between DHLists and Top 10k.}
  \label{table:coverage-top10k}
  \centering
  \begin{tabular}{lrr}
  \toprule
  Top 10k & DHList-Global & DHList-Japan\\
  \midrule
  Alexa & 224 (1.64\%) & 8 (0.05\%)\\
  Umbrella & 32 (0.23\%) & 1 (0.01\%)\\
  Majestic & 572 (4.19\%) & 17  (0.11\%)\\
  Tranco & 324 (2.38\%) & 8  (0.05\%)\\
  Secrank & 22 (0.16\%) & 1 (0.01\%)\\
  CrUX & 132 (0.97\%) & 9 (0.06\%)\\
  \bottomrule
  \end{tabular}
\end{table}

\begin{table}[!t]
  \centering
  \caption{Overlap between DHLists and Top 1M.}
  \label{table:coverage-top1m}
  \centering
  \begin{tabular}{lrr}
  \toprule
  Top 1M & DHList-Global & DHList-Japan\\
  \midrule
  Alexa & 3,842 (28.2\%) & 459 (3.0\%)\\
  Umbrella & 2,873 (21.1\%) & 89 (0.6\%)\\
  Majestic & 7,296 (53.5\%) & 1,931 (12.8\%)\\
  Tranco & 6,258 (45.9\%) & 1,395 (9.3\%)\\
  Secrank & 1,306 (9.6\%) & 279 (1.9\%)\\
  CrUX & 3,824 (28.0\%) & 1,779 (11.8\%)\\
  \bottomrule
  \end{tabular}
\end{table}

\subsection{Trustworthiness of Domains in DHLists}
\label{sec:trustworthiness-dhlists}
To investigate the trustworthiness of domains in DHLists, we use VirusTotal's commercial version, a prominent industry security service, for an objective comparison of the risks associated with domains in DHList-Global (13,640 domains) and DHList-Japan (15,601 domains) against those in Tranco-Top10k.
VirusTotal utilizes over 80 diverse antivirus software and scanning engines to conduct comprehensive scans of websites and domains.

Table~\ref{table:vt-malicious} presents the number of domain names labeled as malicious by multiple engines using VirusTotal.
VirusTotal can analyze any domain with up to 87 different engines.
On VirusTotal, labels for each engine are standardized; the label ``malicious'' signifies that the engine has identified the domain as harmful.
Table~\ref{table:vt-malicious} displays the number of domains considered malicious when the number of engines with malicious labels exceeds a certain threshold, denoted as \#Engines.
For example, the column \#Engines$\geq 5$ in Table~\ref{table:vt-malicious} shows the number of domains in each list flagged as malicious by five or more engines.

Table~\ref{table:vt-malicious} provides two key insights into VirusTotal's results concerning maliciousness.
Firstly, irrespective of the \#Engines threshold, the total count of malicious domains identified by VirusTotal decreases in the order of Tranco-Top10k, DHList-Global, and DHList-Japan.
Despite Tranco-Top10k having the fewest domains (10,000), the risk associated with domains in the proposed DHLists is lower than that of Tranco-Top10k.
Second, a significant number of domains (e.g., 558, 435, and 150) are identified as malicious across the three lists when \#Engines$\geq 1$.
Since VirusTotal outputs results from multiple engines without specific weighting, engines prone to false positives can disproportionately influence the overall outcomes.
Consequently, many studies only label targets (e.g., domains, URLs, and files) as malicious when multiple engines concur~\cite{DBLP:conf/uss/ZhuSYQZS020}.
Numerous security researchers, ourselves included, have scrutinized VirusTotal's labels and determined that three specific engines account for a majority of the false positives, with domains flagged solely by these engines comprising over 80\% of the false positives when \#Engines=1.
However, for the sake of reproducibility, we have not excluded these results, and the figures presented in this study are compiled directly from VirusTotal's findings.

\begin{table*}[!t]
  \caption{Number of domains flagged as malicious.}
  \label{table:vt-malicious}
  \centering
  \begin{tabular}{lr|rrrrr}
  \toprule
    & \#Listed & \multicolumn{5}{l}{\#VirusTotal Malicious Domains} \\
    &  Domains & \#Engines $\geq 1$ & \#Engines $\geq 2$ & \#Engines $\geq 3$ & \#Engines $\geq 4$ & \#Engines $\geq 5$\\
  \midrule
  Tranco-Top10k & 10,000  & 558  & 106  & 46  & 25  & 16 \\
  DHList-Global & 13,640  & 435  & 39  & 15  & 7  & 5 \\
  DHList-Japan & 15,058  & 150  & 13  & 3  & 2  & 1 \\
  \bottomrule
  \end{tabular}
\end{table*}

\begin{table*}[!t]
  \centering
  \caption{Performance comparison of ML algorithms.}
  \label{table:ml-eval}
  \begin{tabular}{lllrrrr}
  \toprule
  List & Date & Algorighm & AUC & Recall & Precision & F1 \\
  \midrule
  DHList-Global & 2022-11-25 & \cellcolor{yellow!25}\textbf{LightGBM} & \cellcolor{yellow!25}\textbf{93.3\%} & \cellcolor{yellow!25}\textbf{82.7\%} & \cellcolor{yellow!25}\textbf{92.6\%} & \cellcolor{yellow!25}\textbf{87.4\%} \\
  DHList-Global & 2022-11-25 & Random Forest & 87.6\% & 73.1\% & 87.8\% & 79.7\% \\
  DHList-Global & 2022-11-25 & Decision Tree & 81.6\% & 81.6\% & 82.0\% & 81.7\% \\
  \midrule
  DHList-Japan & 2022-11-25 & \cellcolor{yellow!25}\textbf{LightGBM} & \cellcolor{yellow!25}\textbf{90.8\%} & \cellcolor{yellow!25}\textbf{78.0\%} & \cellcolor{yellow!25}\textbf{89.4\%} & \cellcolor{yellow!25}\textbf{83.3\%} \\
  DHList-Japan & 2022-11-25 & Random Forest & 84.9\% & 70.3\% & 82.9\% & 76.1\% \\
  DHList-Japan & 2022-11-25 & Decision Tree & 75.8\% & 77.0\% & 75.2\% & 76.1\% \\
  \bottomrule
  \end{tabular}
\end{table*}

\begin{table*}[!t]
  \caption{Number of domains flagged as malicious after Labeling (❹) and DHList Generation (❻).}
  \label{table:vt-malicious-untrustworthy}
  \centering
  \tabcolsep=1.0mm
  \begin{tabular}{llr|rrrrr}
  \toprule
      &      & \#Untrustworthy & \multicolumn{5}{l}{\#VirusTotal Malicious Domains} \\
  List & Step & Domains & \#Engines $\geq 1$ & \#Engines $\geq 2$ & \#Engines $\geq 3$ & \#Engines $\geq 4$ & \#Engines $\geq 5$\\
  \midrule
  DHList-Global & ❹ Labeling & 1,498  & 169  & 23  & 14  & 9  & 5  \\
  DHList-Global & ❻ DHList Generation & 86,619  & 2,203  & 177  & 91  & 60  & 36  \\
  \midrule
  DHList-Japan & ❹ Labeling & 1,925  & 40  & 6  & 1  & 1  & 1  \\
  DHList-Japan & ❻ DHList Generation & 79,200  & 1,344  & 130  & 68  & 41  & 26  \\
  \bottomrule
  \end{tabular}
\end{table*}

\subsection{Characteristics of ML Models}
\label{sec:characteristics-ml}
We report the characteristics of the ML models used in this study.
Specifically, we present the results of using different ML algorithms within DomainHarvester, the significance of the designed feature dimensions, and the security scan results for the domain names identified as untrustworthy.

\noindent\textbf{ML Algorithm.}
We implement PU Learning with several ML algorithms: Decision Tree, Random Forest, and LightGBM, comparing their accuracies.
As these are all tree-based algorithms, they do not require feature scaling, which is advantageous in environments where data can change weekly, as in this study.
Although all tree-based, there are distinctions among them: Decision Tree is a foundational algorithm, Random Forest an ensemble method utilizing multiple decision trees, and LightGBM a gradient boosting framework that also employs multiple decision trees.
We evaluate their performance using Cross Validation on the Training Dataset of the last week (2022-11-25) for both DHList-Global and DHList-Japan.
We employ metrics such as AUC (Area Under the Curve), Recall, Precision, and F1 score.
AUC measures the area under the ROC (Receiver Operating Characteristic) curve to gauge classifier performance.
Recall quantifies the proportion of true positives correctly identified, whereas Precision measures the accuracy of positive identifications.
The F1 score, the harmonic mean of Recall and Precision, indicates overall classifier performance.
Table~\ref{table:ml-eval} displays the results for each algorithm, highlighting LightGBM's superior performance across all metrics for both DHList-Global and DHList-Japan.
Subsequent results discussed pertain to those obtained using LightGBM.

\noindent\textbf{Feature Importance.}
We assess the feature importance within the ML models.
Given that LightGBM is tree-based, feature importance is calculated by counting how often a feature is used for node splitting across all decision trees.
Table~\ref{table:list-features} shows the importance for each feature for DHList-Global and DHList-Japan.
Two observations emerge: first, the features critical for identifying untrustworthiness vary by seed.
For example, IPv6-related features (No.5 and 6) were not influential in generating DHList-Japan.
Second, Language Model-based features such as Title and Link Text are frequently referenced, indicating that web context features significantly impact the identification of untrustworthy domains.
It's important to note that feature importance values are relative, with higher numbers denoting greater importance, but lower values do not necessarily imply the feature is redundant.

\noindent\textbf{Eliminated Domains.}
We showcase the scanning results for untrustworthy domain names eliminated during the Labeling (❹) and DHList Generation (❻) phases via VirusTotal.
Table~\ref{table:vt-malicious-untrustworthy} lists the domain names deemed malicious when labeled by multiple engines on VirusTotal.
The table also indicates that both labeling and generation steps effectively excluded risky domains.
Our labeling and ML model aim not to pinpoint malicious entities but to flag domains as untrustworthy if they lack the trustworthiness needed for allowlisting or pose any risk.
Our process includes domains without active NS records or those using domain parking, which may not be accurately evaluated by VirusTotal.
Consequently, not all domains are flagged as malicious by VirusTotal.
Since our focus is on unpopular domains, they may not be frequently scanned by VirusTotal, and many were rescanned at our request.

\section{Discussion}
\label{sec:discussion}

\subsection{Ethical Consideration}
\label{sec:ethical}
This study raises no ethical concerns.
As outlined in Section~\ref{sec:proposed-system}, the seed list in DomainHarvester should be resubmitted for crawling no less than one week after the prior submission to mitigate high-frequency access to web servers.
Additionally, when accessing the web in each cycle, the interval between accesses to the same server is controlled to be at least three seconds.
DomainHarvester does not utilize any user or telemetry data that only specific companies possess.
To evaluate the system and the DHLists generated by it in Section~\ref{sec:evaluation}, we employed existing top lists and data from DNSDB and VirusTotal.
This data comprises solely statistical information and results from the respective detection engines.
We did not seek approval from the Institutional Review Board (IRB), as our organization does not consider our research to involve human subjects.

\subsection{Limitation}
\label{sec:limitation}
We discuss the potential for attackers to insert arbitrary domain names into the DHLists from their perspective.
Although we propose a novel method for creating allow lists distinct from top lists, the risk of attackers inserting arbitrary domain names into top lists is well-documented~\cite{DBLP:conf/isw/RweyemamuLWRK19}.
This remains an open challenge for allow lists, a challenge our system does not claim to fully address. DomainHarvester does not rely on external traffic volumes, which can be manipulated by attackers, as inputs, unlike top lists.
Furthermore, it does not accept seed URLs from online third-party submissions.
Instead, seed URLs are selected by the list provider, preventing attackers from inputting arbitrary URLs into the seeds.
Moreover, unlike top lists that produce a list based solely on popularity at a single point in time, DomainHarvester autonomously generates the final DHLists through regular Web and DNS crawls, leveraging Language Models and multiple feature dimensions, and is designed to exclude anything with even slight changes in trustworthiness.
Consequently, the probability of arbitrary domain names being included in the final DHLists is minimal.
Additionally, integrating specific measures into the current DomainHarvester to mitigate this issue is straightforward.
Specifically, as used for evaluation in Section~\ref{sec:trustworthiness-dhlists}, performing regular weekly security scans, such as those by VirusTotal, on the domain names and their underlying URLs that make it to the final DHLists and adding a filtering step to remove any domain name from the DHLists if it poses some risk (such as malware, phishing, or being compromised) is one such measure.
Of course, the scanning method is not limited to VirusTotal; other security scanning services can also be utilized.
However, this filtering step has a trade-off in that it may impair reproducibility in Internet measurement or security research.
This is because commercial threat intelligence often comes with prohibitive costs and licensing restrictions that impede its use~\cite{DBLP:conf/uss/BouwmanGEDKE20}.
While in many cases allow lists have been used blindly without additional security scans, there should always be a discussion about the option of performing such security scans and filtering, whether the allow list is based on top lists or DHLists.

\section{Conclusion}
\label{sec:conclusion}
This study highlighted the neglect of less popular, yet trustworthy domains when using top lists as the primary source for constructing allow lists in web security.
We introduced DomainHarvester, a system that reveals these neglected domains, demonstrating that our DHLists include a wider array of such domains compared to top lists and significantly diminish the inclusion of genuinely malicious ones.
This approach could potentially minimize false positives, benefiting both users and security analysts.

This study is the first to tackle the issue of allow lists relying solely on popularity, and we encourage further research that explores this topic from diverse angles or examines region-specific characteristics.
Such efforts would aid in ensuring that the multifaceted web landscapes of non-English-speaking regions are not inaccurately labeled as malicious.

\bibliographystyle{IEEEtran}
\bibliography{main}

\end{document}